\documentstyle[aps,prd,preprint,eqsecnum]{revtex}

\newcommand{\be}{\begin{equation}}
\newcommand{\ee}{\end{equation}}
\newcommand{\ba}{\begin{eqnarray}}
\newcommand{\ea}{\end{eqnarray}}
\newcommand{\bas}{\begin{eqnarray*}}
\newcommand{\eas}{\end{eqnarray*}}
\newcommand{\hsp}{\hspace{.5cm}}

\newcommand{\spaceand}{\hsp {\rm and} \hsp}

\tightenlines

\begin{document}

\draft

\title{Gravitational Entropy and Global Structure}
\author{S.W. Hawking\thanks{email:  S.W.Hawking@damtp.cam.ac.uk} and 
        C.J. Hunter\thanks{email:  C.J.Hunter@damtp.cam.ac.uk}}
\address{Department of Applied Mathematics and
      Theoretical Physics, University of Cambridge,
      \\Silver Street, Cambridge CB3 9EW, United Kingdom
       }
\date{5 August 1998}

\maketitle

\begin{abstract}
 The underlying reason for the existence of gravitational entropy is 
traced to the impossibility of foliating topologically non-trivial 
Euclidean spacetimes with a time function to give a unitary Hamiltonian 
evolution. In $d$ dimensions the entropy can be expressed in terms of the 
$d-2$ obstructions to foliation, bolts and Misner strings, by a universal 
formula. We illustrate with a number of examples including spaces with nut 
charge. In these cases, the entropy is not just a quarter the area of the bolt, 
as it is for black holes. 
\end{abstract}


\narrowtext

\section{Introduction}
\label{sec:Introduction}

The first indication that gravitational fields could have entropy came when 
investigations \cite{Christodoulou70} of the Penrose process for extracting 
energy from a 
Kerr black hole showed that there was a quantity called the irreducible mass 
which could go up or stay constant, but which could never go down. Further work 
\cite{Hawking71} showed that this irreducible mass was proportional to the 
area of the 
horizon of the black hole and that the area could never decrease in the classical theory, even in 
situations where black holes collided and merged together. There was an obvious 
analogy with the Second Law of Thermodynamics, and indeed black holes were 
found 
to obey analogues of the other laws of Thermodynamics as well 
\cite{BardeenCarterHawking73}.  But
it was Jacob Bekenstein who took the bold step \cite{Bekenstein73} of 
suggesting the area 
actually was the physical entropy, and that it counted the internal states of 
the black hole. The inconsistencies in this proposal were removed when it was 
discovered that quantum effects would cause a black hole to radiate like a hot body 
\cite{Hawking74,Hawking75}. 

For years people tried to identify the internal states of black holes in terms 
of fluctuations of the horizon. Success seemed to come with the paper of 
Strominger and Vafa \cite{StromingerVafa96} which was followed by a host of 
others. However, 
in light of recent work on anti-de Sitter space 
\cite{AntideSitter}, one could reinterpret these 
papers as establishing a relation between the entropy of the black hole and the 
entropy of a conformal field theory on the boundary of a related anti-de Sitter 
space. This work, however, left obscure the deep reason for the existence of 
gravitational entropy. In this paper we trace it to the fact that general 
relativity and its supergravity extensions allow spacetime to have more than one
topology for given boundary conditions at infinity. By topology, we mean 
topology in the Euclidean regime. The topology of a Lorentzian spacetime can 
change with time only if there is some pathology, such as a singularity, or 
closed time-like curves. In either of these cases, one would expect the theory 
to break down.

The basic premise of quantum theory is that time translations are unitary 
transformations generated by the Hamiltonian. In gravitational theories the 
Hamiltonian is given by a volume integral over a hypersurface of constant time,
plus surface integrals at the boundaries of the hypersurface. The volume 
integral vanishes if the constraints are satisfied, so the numerical value of 
the Hamiltonian comes from the surface terms. 
However, this does not mean that the 
energy and momentum reside on these boundaries. Rather it reflects that these 
are global quantities which cannot be localized. We shall argue the same is 
true of entropy: it is a global property and cannot be localised as 
horizon states.

If the spacetime can be foliated by a family of surfaces of constant time, the 
Hamiltonian will indeed generate unitary transformations and there will be no 
gravitational entropy. However, if the topology of the Euclidean spacetime is 
non-trivial, it may not be possible to foliate it by surfaces that do not 
intersect 
each other and which agree with the usual Euclidean time at infinity. In this 
situation, the concept of unitary Hamiltonian evolution breaks down and mixed 
states with entropy will arise. We shall relate this entropy to the obstructions
to foliation. It turns out that that the entropy of a $d$ dimensional Euclidean 
spacetime ($d>2$) can expressed in terms of bolts ($d-2$ dimensional 
fixed point sets of the
time translation Killing vector) and Misner strings (Dirac strings in the Kaluza
Klein reduction with respect to the time translation Killing vector) by the 
universal formula:
\begin{equation} 
S ={1\over 4G}({\cal A}_{\rm bolts}+{\cal A}_{\rm MS})-\beta H_{\rm MS},
\end{equation} 
where $G$ is the $d$ dimensional Newton's constant, ${\cal A}_{\rm bolts}$ and 
${\cal A}_{\rm MS}$ are respectively the $d-2$ volumes in the Einstein frame 
of the bolts 
and Misner strings and $H_{MS}$ is the Hamiltonian surface term on the Misner 
strings. Where necessary, subtractions should be made for the same quantities in
a reference background which acts as the vacuum for that sector of the theory.

The plan of this paper is as follows. In section \ref{sec:Hamiltonian}
 we describe the ADM formalism
and the expression for the Hamiltonian in terms of volume and surface integrals.
In section \ref{sec:Thermodynamics} we introduce thermal ensembles and give an 
expression for the 
action and entropy of Euclidean metrics with a $U(1)$ isometry group. This is
illustrated in section \ref{sec:Examples} by some examples.  In section
\ref{sec:Conclusions} we draw some morals. 

\section{Hamiltonian}
\label{sec:Hamiltonian}

Let $\bar{\cal M}$ be a $d$-dimensional Riemannian manifold with metric 
$g_{\mu\nu}$ and covariant derivative $\nabla_\mu$, which has an 
imaginary time coordinate $\tau$ that foliates 
$\bar{\cal M}$ into non-singular hypersurfaces $\{ \Sigma_\tau \}$ of constant 
$\tau$.  The metric and covariant derivative on $\Sigma_\tau$ are $h_{ij}$ and
$D_i$.  If $\bar{\cal M}$ is non-compact then it will have a boundary 
$\partial \bar{\cal M}$, 
which can include internal components as well as a boundary at infinity.  The
$d-2$ dimensional surfaces, $B_\tau = \partial \bar{\cal M} \cap \Sigma_\tau$, 
are the boundaries of the hypersurfaces $\Sigma_\tau$ and a foliation of 
$\partial \bar{\cal M}$.  We will use Greek letters to denote indices on 
$\bar{\cal M}$, and roman letters for indices on $\Sigma_\tau$. 
 
The Euclidean action for a gravitational field coupled to both a Maxwell 
and $N$ general matter fields is   
\be
 I = -\frac{1}{16\pi G} \int_{\cal M} d^dx\sqrt{g}\, [R - F^2 + 
     {\cal L}(g_{\mu\nu},\phi^A)] - 
    \frac{1}{8\pi G} \int_{\cal M} d^{d-1}x \sqrt{b}\, \Theta(b),
\ee
where $R$ is the Ricci scalar, $F_{\mu\nu}$ is the Maxwell field tensor, and
${\cal L}(g_{\mu\nu},\phi^A)$ is an arbitrary Lagrangian for the  
fields $\phi^A$ (A=1..N), where any tensor indices for $\phi^A$ are suppressed.
We assume the $\cal L$ contains only first derivatives, and hence does not need
an associated boundary term.  

In order to perform the Hamiltonian decomposition of the action, we write   
metric in ADM form \cite{ArnowittDeserMisner62},
\be
 ds^2 = N^2d\tau^2 + h_{ij}(dx^i+N^id\tau)(dx^j+N^jd\tau).
\ee
This defines the lapse function $N$, the shift vector $N^i$, and the induced 
metric on $\Sigma_\tau$, $h_{ij}$.  
We can rewrite the action (see \cite{HawkingHorowitz96,HawkingHunter96} for 
details) as
\begin{equation}
 I = \int d\tau \left[ \int_{\Sigma_\tau} d^{d-1}x (P^{ij}\dot{h}_{ij} + 
          E^i\dot{A}_i + \sum_{A=1}^N \pi^A \dot{\phi}^A ) + H\right], 
\end{equation}
where $P^{ij}$, $E^i$ and $\pi^A$ are the momenta conjugate to the dynamical
variables $h_{ij}$, $A_i$ and $\phi^A$ respectively.  The Hamiltonian, $H$, 
consists of a volume integral over $\Sigma_\tau$, and a boundary
integral over $B_\tau$.  

The volume term is 
\be
 H_c = \int_{\Sigma_\tau} d^{d-1}x\,\left[N{\cal H} + N^i{\cal H}_i + 
       A_0(D_i E^i - \rho) + \sum_{A=1}^M \lambda^A C^A \right],
\ee
where $N$, $N^i$, $A_0$ and $\lambda^A$ are all Lagrange multipliers for the
constraint terms $\cal H$, ${\cal H}_i$, $D_i E^i - \rho$ and $C^A$.  The number
of constraints, $M$,  which arise from the matter Lagrangian depends on its 
exact form.  $\rho$ is the electromagnetic charge density.
Since the constraints all vanish on metrics that satisfy the field equations, 
the volume term makes no contribution to the Hamiltonian when it is evaluated 
on a solution.

The boundary term is
\be
 \label{eqn:Hamiltonian_b}
 H_b = -\frac{1}{8\pi G} \int_{B_\tau} \sqrt{\sigma}[ Nk + 
         u_i(K^{ij}-Kh^{ij})N_j + 2 A_0 F^{0i} u_i + f(N,N^i,h_{ij},\phi^A) ],
\ee
where $\sqrt{\sigma}$ is the area element of $B_\tau$, 
$k$ is the trace of the second fundamental form of $B_\tau$ as embedded in
$\Sigma_\tau$, 
$u_i$ is the outward pointing unit normal to $B_\tau$, 
$K_{ij}$ is the
second fundamental form of $\Sigma_\tau$ in $\bar{\cal M}$, 
and $f(N,N^i,h_{ij},\phi^A)$ is some function which depends on the 
form of the matter Lagrangian.  

Generally the surface term will make both the action and the Hamiltonian 
infinite. In order to obtain a finite result, it is sensible to consider 
the difference between the 
action or Hamiltonian, and those of some reference background solution.  We 
pick the background such that the solution approaches it at infinity 
sufficiently rapidly so that the difference in the action and Hamiltonian are
well-defined and finite. 
This reference background acts as the vacuum 
for that sector of the quantum theory. It is normally taken to be flat space or
anti-de Sitter space, but we will consider other possibilities.  We will 
denote background quantities with a tilde, although in the interest of clarity,
they will be omitted for most calculations.

\section{Thermodynamic  Ensembles}
\label{sec:Thermodynamics}

In order to discuss quantities like entropy, one defines the partition function 
for an ensemble  with temperature $T =\beta ^{-1}$, angular velocity $\Omega $ 
and electrostatic potential $\Phi $ as:
\begin{equation}
 {\cal Z} ={\rm Tr}\, e^{-\beta (E+\Omega \cdot J +\Phi Q)}
= \int D[g] D[\phi ] e^{-I [g,\phi ]},
\end{equation}
 where the path integral is taken over all metrics and fields that agree with 
the reference background at infinity and are periodic under the combination of 
a Euclidean time translation $\beta $, a rotation through an angle 
$\beta \Omega $ and a gauge transformation $\beta \Phi $. 
The partition function includes 
factors for electric-type charges  such as mass, angular momentum and electric 
charge, but not for magnetic-type charges such as nut charge and magnetic 
charge. This is because the boundary conditions of specifying the metric and 
gauge 
potential on a $d-1$ dimensional 
surface at infinity do not fix the electric-type charges. 
Each field configuration in the path integral therefore has to be weighted with 
the appropriate factor of the exponential of minus charge times the 
corresponding thermodynamic potential. Magnetic-type charges, on the other hand,
are fixed by the boundary conditions and are the same for all field 
configurations in the path integral. It is therefore not necessary to include 
weighting factors for magnetic-type charges in the partition function. 

The lowest order contribution to the partition function will be
\begin{equation}
 {\cal Z} =\sum  e^{-I},
\end{equation}
where $I $ are the actions of Euclidean solutions with the given boundary 
conditions. The reference background, periodically identified,
will always be one 
such solution and, by definition, it will have zero action. However, we shall be
concerned in this paper with situations where there are additional Euclidean 
solutions with different topology which also have a $U(1)$ isometry group that 
agrees with the periodic identification at infinity. This includes not only
black holes 
and p-branes, but also more general classes of solution, as we shall show in the
next section. 

In $d$ dimensions the
Killing vector $K={\partial /\partial \tau}$ will have zeroes on surfaces of 
even codimension 
which will be fixed points of the isometry group. The $d-2$ dimensional
fixed points sets 
will play an important role. We shall generalise the terminology of 
\cite{GibbonsHawking79,Dowker96,Hunter98} and call them bolts.

Let $\tau$ with period $\beta $ be the parameter of the $U(1)$ 
isometry group. Then the
metric can be written in the Kaluza Klein form:
\begin{equation}
 ds^2 = \exp\left[-\frac{4\sigma}{\sqrt{d-2}}\right](d\tau + \omega_i dx^i)^2 + 
     \exp\left[\frac{4\sigma}{(d-3)\sqrt{d-2}}\right] \gamma_{ij}dx^idx^j,
\end{equation}
where $\sigma $, $\omega 
_i$ and $\gamma _{ij}$ are fields on the space $\Xi$ of orbits of the isometry 
group. $\Xi$ would be singular at the fixed point so one has to leave them out  
and introduce $d-2$ boundaries to $\Xi$. 

The coordinate $\tau $ can be changed 
by a Kaluza-Klein gauge transformation:
\begin{equation}
 \label{eqn:gauge}
\tau ^{\prime}=\tau +\lambda,
\end{equation}
 where $\lambda $ is a function on $\Xi$. This changes the one-form $\omega$ by 
$d \lambda$ but leaves the field strength $F = d \omega$ unchanged. 
If the orbit 
space $\Xi$ has non-trivial homology in dimension two, then the two-form $F$ 
can have 
non-zero integrals over two-cycles in $\Xi$. In this case, the one-form 
potential 
$\omega$ will have Dirac-like string singularities on surfaces of dimension 
$d-3$ in $\Xi$. The foliation of the spacetime by surfaces of constant $\tau$ 
will break down at the fixed points of the isometry. It will also break down on 
the string singularities of $\omega$ which we call Misner strings, after 
Charles 
Misner who first realized their nature in the Taub-NUT solution \cite{Misner63}.
Misner strings are surfaces of dimension $d-2$ in the spacetime $\cal M$.

 In order to do a Hamiltonian treatment using surfaces of constant $\tau $, one 
has to cut out small neighbourhoods of the fixed point sets and of any Misner 
strings leaving a manifold $\bar{\cal M}$. On $\bar{\cal M}$ one has the usual 
relation between the action and Hamiltonian:
\begin{equation}
 I = \int d\tau \left[ \int_{\Sigma_\tau} d^{d-1}x (P^{ij}\dot{h}{ij} + 
          E^i\dot{A}_i + \sum_A \pi^A \dot{\phi}^A ) + H \right] 
\end{equation}
Because of the $U(1)$ isometry, the time derivatives will all be zero. Thus the 
action of $\bar{\cal M}$ will be
\begin{equation}
 \label{eqn:Mbar_action}
 I(\bar{\cal M}) = \beta H
\end{equation}
To get the action of the whole spacetime $\cal M$, one now has to put back the 
small 
neighbourhoods of the fixed point sets and the Misner strings that were cut out. 
In the limit that the neighbourhoods shrink to zero, their volume contributions 
to the action will be zero. However, the surface term associated with the 
Einstein Hilbert action will give a contribution to the action of $\cal M$ of
\begin{equation} 
 I({\cal M} - \bar{\cal M}) = 
      -\frac{1}{4 G}( {\cal A}_{\rm bolts}+{\cal A}_{\rm MS} ),
\end{equation}
where ${\cal A}_{\rm bolts}$ and ${\cal A}_{\rm MS}$ are respectively the total
area of the bolts and the Misner strings in the spacetime.
The contribution of the Einstein Hilbert term to the action from lower 
dimensional fixed points will be zero. The contribution at bolts and Misner 
strings from higher order curvature terms in the action will be small in the 
large area limit.

The Hamiltonian in (\ref{eqn:Mbar_action})
 will come entirely from the surface terms. In a 
topologically trivial spacetime, the surfaces of $\tau $ will have boundaries 
only at infinity. However, in more complicated situations, the surfaces will 
also have boundaries at the fixed point sets and Misner strings. 
The Hamiltonian 
surface terms at the fixed points will be zero because the lapse and shift 
vanish there. On the other hand, although the lapse is zero, the shift won't 
vanish on a Misner string. Thus there will be a Hamiltonian surface term on a 
Misner string given by the shift times a component of the second fundamental 
form of the constant $\tau$  surfaces. The action of $\cal M$ is therefore
\begin{equation}
 \label{eqn:action_thermo}
 I({\cal M}) = \beta (H_{\infty}+H_{\rm MS})
         -\frac{1}{4 G}({\cal A}_{\rm bolts}+{\cal A}_{\rm MS}).
\end{equation}
On the other hand, by thermodynamics:
\begin{equation}
 \log Z = S -\beta (E +\Omega \cdot J + \Phi Q).
\end{equation}
But,
\begin{equation}
H_{\infty}= E +\Omega \cdot J + \Phi Q,
\end{equation}
so
\begin{equation} 
 \label{eqn:entropy}
S = \frac{1}{4G}({\cal A}_{\rm bolts}+{\cal A}_{\rm MS})-\beta H_{\rm MS}.
\end{equation} 
The areas and Misner string Hamiltonian in equation (\ref{eqn:entropy}) are to
be understood as differences from the reference background.

In order for the thermodynamics to be sensible, it must be invariant under
the gauge transformation (\ref{eqn:gauge}) which rotates the imaginary time
coordinate.
Because the action (\ref{eqn:action_thermo}) is gauge invariant, we see that
the entropy will also be, provided that $H_\infty$ is independent of the gauge. 
In appendix \ref{app:Infinity}, we show that $H_\infty$ is indeed gauge 
invariant, and hence the entropy is well-defined, for metrics satisfying 
asymptotically flat (AF), asymptotically
locally flat (ALF) or asymptotically locally Euclidean (ALE) boundary  
conditions.

Previous expositions of gravitational entropy have not included ALF and ALE
metrics.  This is presumably because these metrics contain Misner strings, and 
hence do not obey the simple ``quarter-area law'', but rather the more 
complicated expression (\ref{eqn:entropy}).   

\section{Examples}
\label{sec:Examples}

In this section we calculate the entropy of some four and five dimensional
spacetimes.  We set $G=1$.
The first example considers the Taub-NUT and Taub-Bolt metrics, which are ALF.  
We then move to solutions of Einstein-Maxwell theory, 
the Israel-Wilson metrics, and calculate the entropy in both the AF and ALF 
sectors.  The Eguchi-Hanson instanton then provides us with an ALE example.
Finally, we calculate the
entropy of $S^5$ for two different $U(1)$ isometry groups, one with a bolt, and
the other with no fixed points but a Misner string, obtaining the same result 
both ways.
The action calculations, reference backgrounds and matching conditions for 
Taub-NUT, Taub-Bolt and Eguchi-Hanson are all presented 
in \cite{Hunter98} and will not be repeated here.


\subsection{Taub-NUT and Taub-Bolt}
\label{sec:Taub}

ALF solutions have a Nut charge, or magnetic type mass, N, as well as the 
ordinary electric type mass, M.  The Nut charge is $\beta C_1/8 \pi$, where 
$C_1$ is the first Chern number of the $U(1)$ bundle
over the sphere at infinity, in the orbit space $\Xi$. 
If the Chern number is zero, then the boundary at infinity is $S^1\times S^2$ 
and the spacetime is AF.  The black hole metrics are saddle points in the path 
integral for the partition function.  They have a bolt on the horizon but no 
Misner strings, and hence  
equation (\ref{eqn:entropy}) gives the usual result for the entropy.  
However, if the Chern number is nonzero, the boundary at infinity is a squashed
$S^3$, and the metric cannot be analytically continued to a Lorentzian metric.
Nevertheless, one can formally interpret the path integral over all metrics
with these boundary conditions as giving the partition function for an ensemble
with a fixed value of the nut charge or magnetic-type mass.

The simplest example of an ALF metric is the  
Taub-NUT instanton \cite{Hawking77}, given by the metric
\begin{equation}
  \label{eqn:Taub_metric}
 ds^2 = V(r)(d\tau + 2N\cos\theta d\phi)^2 + \frac{1}{V(r)}dr^2 + (r^2-N^2)
  (d\theta^2 + \sin^2\theta d\phi^2),
\end{equation}
where $V(r)$ is
\begin{equation}
 V_{TN}(r) = \frac{r-N}{r+N}.
\end{equation}
In order to make the solution regular, we consider the region
$r \geq N$ and let the period of $\tau$ be $8\pi N$.
The metric has a nut at $r=N$, with a Misner string running along the 
$z$-axis from the nut out to infinity.  

The Taub-Bolt instanton \cite{Page78} is also given by the metric
(\ref{eqn:Taub_metric}).  However, the function $V(r)$ is different,
\begin{equation}
 V_{TB}(r) = \frac{(r-2N)(r-N/2)}{r^2-N^2}.
\end{equation}
The solution is regular if we consider the
region $r \geq 2N$ and let $\tau$ have period $\beta = 8\pi N$.
Asymptotically, the Taub-Bolt instanton is also ALF.
There is a bolt of area $12\pi N^2$ at $r=2N$ which is a source for a Misner
string along the $z$-axis.

In order to calculate the Hamiltonian of the Taub-Bolt instanton, we need to
use a scaled Taub-NUT metric as the reference background.
We can then calculate the Hamiltonian at infinity,
\begin{equation}
 H_\infty  =  \frac{N}{4},
\end{equation}
and the contribution from the boundary around the Misner string,  
\begin{equation}
 H_{MS} = -\frac{N}{8}.
\end{equation}
The area of the Misner string is $-12\pi N^2$ (that is, the area of the Misner
string is greater in the Taub-NUT background than in Taub-Bolt).
Combining the Hamiltonian, Misner string and bolt
contributions yields an action and entropy of
\begin{equation}
 I = \pi N^2 \spaceand S = \pi N^2.
\end{equation}

It would be interesting to relate this
entropy to the entropy of a conformal field theory defined on the boundary of 
the spacetime. This may be possible by considering Euclidean Taub-NUT 
anti-de Sitter, and other spacetimes asymptotic to it. The boundary at infinity
is a squashed three sphere, and the squashing tends to a constant 
at infinity. One would then compare the entropy of asymptotically Taub-NUT 
anti-de Sitter spaces with the partition function of a conformal field theory 
on the squashed three sphere.  Work on this is in progress \cite{Hawking99}. 

\subsection{Israel-Wilson}
\label{sec:Israel-Wilson}

 The Euclidean Israel-Wilson family of metrics \cite{Whitt85,Yuille87}
are solutions of the Einstein-Maxwell equations with line element
\be
 ds^2 = \frac{1}{UW}(d\tau + \omega_idx^i)^2 + UW \gamma_{ij}
  dx^i dx^j,
\ee
where $\gamma_{ij}$ is a flat three-metric and $U$, $W$ and $\omega_i$ are
real-valued functions.
The electromagnetic field strength is
\be
 F = \partial_i \Phi (d\tau + \omega_j dx^j)\wedge dx^i + UW\!\sqrt{\gamma}\,
        \epsilon_{ijk}\gamma^{kl}\partial_l \chi \, dx^i \wedge dx^j,
\ee
with complex potentials $\Phi$ and $\chi$ given by
\ba
 \Phi & = & \frac{1}{2}\left\{ \left(\frac{1}{U}-\frac{1}{W}\right)\cos\alpha
  + \left(\frac{1}{U}+\frac{1}{W}\right)i\sin\alpha \right\} \hsp {\rm and} \\
 \chi & = & -\frac{1}{2}\left\{ \left(\frac{1}{U}+\frac{1}{W}\right)\cos\alpha
  + \left(\frac{1}{U}-\frac{1}{W}\right)i\sin\alpha \right\}.
\ea
For $F^2$ to be real, we need to take $\Phi$ and $\chi$
to be either entirely real or purely imaginary.  Taking them to be real,
we obtain the magnetic solution,
\be
 \Phi_{\rm mag} = \frac{1}{2}\left(\frac{1}{U}-\frac{1}{W}\right)
             \spaceand
 \chi_{\rm mag}  = -\frac{1}{2}\left(\frac{1}{U}+\frac{1}{W}\right).
\ee
The dual of the magnetic solution is the electric one, with imaginary
potentials.  Calculating the square of the field strengths,
\be
 F_{\rm mag}^2 = (D U^{-1})^2 + (D W^{-1})^2 = - F_{\rm elec}^2.
\ee
We consider only the magnetic solutions here.  The action and entropy
calculations for the electric case are similar.

$U$, $W$ and $\omega_i$
are determined by the equations
\be
 D_i D^i U = 0 = D_i D^i W \spaceand \frac{1}{\sqrt{\gamma}}\gamma_{ij}
    \epsilon^{jkl} \partial_k \omega_l = W D_i U - U D_i W,
\ee
where $D_i$ is the covariant derivative for $\gamma_{ij}$.  The solutions for
$U$ and $W$ are simply three-dimensional harmonic functions, and we will take
them to be of the form
\be
 U = 1 + \sum_{I=1}^N \frac{a_I}{|x-y_I|} \spaceand
 W = 1 + \sum_{J=1}^M \frac{b_J}{|x-z_J|},
\ee
where $y_I$ and $z_J$ are called the mass and anti-mass points
respectively, and comprise the fixed point set of $\partial_\tau$.  We assume 
that the points have positive mass, i.e., $a_I,b_J > 0$.

There will generically be conical singularities in the metric at the mass and
anti-mass points.  In order to remove them we must apply the constraint
equations,
\be
 U(z_J)b_J = \frac{\beta}{4\pi} = W(y_I)a_I,
\ee
where $\beta$ is the periodicity of $\tau$.  Note that these equations
hold for each value of $I$ and $J$, i.e., no summation is implied.
While the resulting spacetime is non-singular,
emanating from each fixed point there will be Misner string
singularities in the metric, and Dirac string singularities in the
electromagnetic potential.  These string singularities will end on either
another fixed point or at infinity.

  The Einstein-Maxwell action is
\be
 I = -\frac{1}{16\pi} \int_{\cal M} d^4x \sqrt{g}\, (R - F^2) -
        \frac{1}{8\pi} \int_{\partial \cal M} d^3x\sqrt{b}\,\Theta(b).
\ee
which we can divide up into a gravitational (Einstein-Hilbert) and an
electromagnetic term, $I = I^{\rm EH} + I^{\rm EM}$.

Since the Ricci scalar, $R$, is zero, the gravitational contribution to the
action is entirely from the the surface term at infinity,
\be
 I^{\rm EH} = -\frac{1}{8\pi} \int_{\partial \cal M} d^3x\sqrt{b}\,\Theta(b).
\ee
Substituting in the metric, we can evaluate this on a hypersurface of radius
$r$,
\be
 \label{eqn:action_EH}
 I^{\rm EH} = -\beta r -
    \frac{\beta}{16\pi} \int_{\partial \Xi} d^2x \sqrt{\sigma}\,
         \frac{u^i D_i(UW)}{UW},
\ee
where $\sigma_{ij}$ is the metric induced on the boundary from $\gamma_{ij}$,
and $u^i$ is the unit normal to the boundary.

We can write the electromagnetic contribution to the action integral as
\ba
 I^{\rm EM} = \frac{1}{16\pi} \int_{\cal M} d^4x\sqrt{g}\, F^2  & = &
  \frac{\beta}{32\pi}\int_\Xi d^3x\sqrt{\gamma}\left[ \frac{D_i D^i W}{U} +
     \frac{D_i D^i U}{W} \right] \nonumber - \\
  & &  \frac{\beta}{32\pi}\int_{\partial \Xi} d^2x \sqrt{\sigma} u^i D_i (UW)
     \left[ \frac{1}{U^2} + \frac{1}{W^2} \right],
\ea
where $\partial \Xi$ is the boundary of $\Xi$ at infinity (since the internal
boundaries about the fixed points will make no contribution).
We can evaluate the volume integral by using the delta function behaviour
of the Laplacians of $U$ and $W$,
\be
 \label{eqn:action_EM}
 I^{\rm EM}   =
       - \frac{\pi}{2}\left(\sum_{I=1}^N a^2_I + 
       \sum_{J=1}^M b_J^2 \right) -
     \frac{\beta}{32\pi}\int_{\partial \Xi} d^2x \sqrt{\sigma} u^i D_i (UW)
  \left[ \frac{1}{U^2} + \frac{1}{W^2} \right].
\ee
Note that the sum is only over mass and anti-mass points which are not
coincident.

  Suppose that we consider metrics with an equal number of nuts and anti-nuts,
\be
 U = 1 + \sum_{I=1}^N \frac{a_I}{|x-y_I|} \spaceand
 V = 1 + \sum_{I=1}^N \frac{b_I}{|x-z_I|}.
\ee
Applying the constraint equations, we see that
\be
 \sum_{I=1}^N a_I = \sum_{I=1}^N b_I \equiv A.
\ee
Hence, the scalar functions asymptotically look like
\be
 U \sim 1 + \frac{A}{r} + {\cal O}(r^{-2}) \spaceand
 W \sim 1 + \frac{A}{r} + {\cal O}(r^{-2}),
\ee
while the vector potential vanishes,
\be
 \omega_i \sim {\cal O}(r^{-2}).
\ee
Thus, at large radius the metric is
\be
 ds^2 \sim \left(1-\frac{2A}{r} \right)d\tau^2 + \left(1+\frac{2A}{r}\right)
     d{\cal E}_3^2,
\ee
so that the boundary at infinity is $S^1 \times S^2$, and the metric is AF.

The background is simply flat space which is scaled so that it matches the
Israel-Wilson metric on a hypersurface of constant radius $R$,
\be
 d\tilde{s}^2 = \left(1-\frac{2A}{R} \right)d\tau^2 +
                    \left(1+\frac{2A}{R}\right) d{\cal E}_3^2,
\ee
and has the same period for $\tau$.
There is no background electromagnetic field.

Using formula (\ref{eqn:action_EH}) for the gravitational contribution to the
action, we obtain, after subtracting off the background term,
\be
 I^{\rm EH} = \frac{\beta}{2} A.
\ee
From equation (\ref{eqn:action_EM}) for the electromagnetic action we get
\be
 I^{\rm EM} = - \frac{\pi}{2}\sum_{I=1}^N (a_I^2 +b_I^2) + \frac{\beta}{2}A.
\ee
Note that the constraint equations imply that $I^{\rm EM}$ is positive.
The total action is therefore positive, and given by 
\be
 I =  \beta A - \frac{\pi}{2} \sum_{I=1}^N (a_I^2 + b_I^2).
\ee

  We can calculate the  Hamiltonian by integrating (\ref{eqn:Hamiltonian_b})
over the boundaries at infinity and around the Misner strings
(note that in the background space there are no Misner strings).
The gravitational contribution from infinity is
\be
 H_\infty = A,
\ee
while the electromagnetic contribution from infinity is zero, because there is
no electric  charge.
On the boundary around the Misner strings, the Hamiltonian is
\be
 H_{\rm MS} = \frac{R}{4} - \frac{\pi}{2\beta} \sum_{I=1}^N(a_I^2+b_I^2),
\ee
where $R$ is the total length of the Misner string.  The area of the Misner
strings is thus
\be
 {\cal A} = \beta R.
\ee
Hence we see that the entropy is
\be
 S = \frac{\pi}{2} \sum_{I=1}^N(a_I^2+b_I^2).
\ee

It is interesting to note that the $N=1$ case is in fact the charged Kerr 
metric subject to the constraint $\beta \Omega = 2\pi$.  This condition 
implies that, unlike the generic Kerr solution, the time translation orbits are
closed.  In a purely bosonic theory this means that the Kerr metric with
$\beta\Omega = 2\pi$ contributes to the partition function,
\be
 {\cal Z} = {\rm tr}\, e^{-\beta H},
\ee
for a non-rotating ensemble.  
However, the partition function will now not contain the factor
$\exp(-\beta\Omega\!\cdot\!J)$.  This means that the entropy will be less than
quarter the area of the horizon by $2\pi J$.  The path integral for the partition
function will also have saddle points at two Reissner-Nordstrom solutions, one
extreme and the other non-extreme.  Both will have the same magnetic charge.
The non-extreme solution will have the same $\beta$ while the extreme one can
be identified with period $\beta$.  The actions will obey
\be
 I_{\rm extreme} > I_{\rm Kerr} > I_{\rm non-extreme}.
\ee
Thus, the non-extreme Reissner-Nordstrom will dominate the partition function.

The situation is different, however, if one takes fermions into account.  In 
this case, the rotation through $\beta \Omega = 2\pi$ changes the sign of the
fermion fields.  This is in addition to the normal reversal of fermions fields
under time translation $\beta$.  Thus, fermions in charged Kerr with $\beta
\Omega = 2\pi$ are periodic under the $U(1)$ time translation group at infinity,
rather than anti-periodic as in Reissner-Nordstrom.  This means that the 
charged Kerr solution contributes to the ensemble with partition function
\be
 {\cal Z} = {\rm tr}\, (-1)^F e^{-\beta H}.
\ee
The extreme Reissner-Nordstrom solution identified with the same periodic spin
structure also contributes to this partition function, but it will be 
dominated by the Kerr solution.  On the other hand, 
the non-extreme 
Reissner-Nordstrom contributes to the normal thermal ensemble with 
partition function
\be
 {\cal Z} = {\rm tr}\, e^{-\beta H}.
\ee
 

If we take a solution with $N$ nuts and $M$ anti-nuts, where $K \equiv N-M> 0$,
then the metric asymptotically approaches
\be
 ds^2 \sim \left(1 - \frac{A+B}{r} \right) \left[d\tau + (A-B)\cos\theta d\phi
     \right]^2 + \left(1+\frac{A+B}{r} \right) [dr^2 + r^2d\Omega_2^2],
\ee
where
\be
 A = \sum_{I=1}^N a_I \spaceand B = \sum_{J=1}^M b_J.
\ee
Applying the constraint equations, we see that
\be
 A-B = K \frac{\beta}{4\pi},
\ee
where $K=M-N > 0$.
Thus, the boundary at infinity will have the topology of a lens space with
$K$ points identified, and hence the metric is ALF.

  If we take $\Phi$ and $\chi$ to be real, then the Maxwell field will also
be real, and will now have both electric and magnetic components.
The choice of gauge 
is then quite important, as it affects how the electromagnetic Hamiltonian
is split between the boundary at infinity and the boundary around the Misner
strings.  We can fix the gauge by requiring the potential to be non-singular on
the boundary at infinity.  Asymptotically, the field is
\be
 A_\mu dx^\mu \sim \left[A_\tau^\infty -\frac{A-B}{2r} \right]d\tau + 
            \left[A_\phi^\infty +\frac{1}{2}(A+B)\right]\cos\theta d\phi,
\ee
where $A_\tau^\infty$ and $A_\phi^\infty$ are the gauge dependent terms that
we have to fix.
By writing the potential in terms of an orthonormal basis, we see that in order
to avoid a singularity we must set 
\be
 A_\tau^\infty = \frac{A+B}{2(A-B)} \spaceand A_\phi^\infty = 0.
\ee  

We can take the background metric to be the multi-Taub-NUT metric 
\cite{GibbonsHawking78} with
$K$ nuts.  This will have the same boundary topology
as the Israel-Wilson ALF solution, and has the asymptotic metric
\be
 ds^2 \sim \left(1-\frac{2NK}{r}\right)[d\tau + 2NK\cos\theta d\phi]^2
   + \left( 1 + \frac{2NK}{r} \right) d{\cal E}_3^2,
\ee
where the periodicity of $\tau$ is $8\pi N$.  By scaling the radial coordinate
and defining the nut charge of each nut, $N$, appropriately, we can match this
to the Israel-Wilson ALF metric on a hypersurface of constant radius $R$.  The
metric is then
\be
 ds^2 \sim \left(1-\frac{2B}{R} -\frac{A-B}{r}\right)
    [d\tau + (A-B)\cos\theta d\phi]^2
   + \left( 1 + \frac{2B}{R} + \frac{A-B}{r} \right) d{\cal E}_3^2,
\ee
where the periodicity of $\tau$ is $\beta$.

Calculating the action, we find that the Einstein-Hilbert contribution is
\be
 I^{\rm EH} = \frac{\beta}{4}(A+B) - \frac{\beta^2}{16\pi}K,
\ee
while the electromagnetic contribution is
\be
 I^{\rm EM} = -\frac{\pi}{2}\left[\sum_{I=1}^N a_I^2 + \sum_{J=1}^M b_J^2\right]
  + \frac{\beta}{4}(A+B).
\ee
Hence the total action is
\be
 I = \frac{\beta}{2}(A+B) - \frac{\beta^2}{16\pi}K
        -\frac{\pi}{2}\left[\sum_{I=1}^N a_I^2 + \sum_{J=1}^M b_J^2\right], 
\ee
which is always positive.

If we calculate the Hamiltonian at infinity, we get
\be
 H_\infty = \frac{3}{4}(A+B) - \frac{\beta}{8\pi}K,
\ee
while the contribution from the Misner string is
\be
 H_{\rm MS} = -\frac{\pi}{2\beta} \left[\sum_{I=1}^N a_I^2 + \sum_{J=1}^M 
       b_J^2\right] - \frac{A+B}{4} + \frac{\beta}{16\pi}K.
\ee
Since the net area of the Misner string is zero, the entropy is simply given 
by the negative of the Misner string Hamiltonian,
\be
 \label{eqn:IW_entropy}
 S = \frac{\pi}{2} \left[\sum_I a_I^2 + \sum_J b_J^2 \right]+  
        \frac{\beta}{4}(A+B) - \frac{\beta^2}{16\pi}K.
\ee
This formula has some strange consequences.  Consider the case of a single nut
and no anti-nuts.  Then the solution is the Taub-NUT instanton with an anti-self
dual Maxwell field on it.  Being self dual, the Maxwell field has zero 
energy-momentum tensor and hence does not affect the geometry, which is 
therefore just that of the reference background.  Yet according to 
equation (\ref{eqn:IW_entropy}), the entropy is $\beta^2/32\pi$.  This entropy
can be traced to the fact that although $A_\mu$ is everywhere regular, the
ADM Hamiltonian decomposition introduces a non-zero Hamiltonian surface term on
the Misner string.  This may indicate that intrinsic entropy is not 
restricted to gravity, but can be possessed by gauge fields as well.  An 
alternative viewpoint would be that the reference background should be  
multi-Taub-NUT with its self dual Maxwell field.  This would change the 
entropy (\ref{eqn:IW_entropy}) to 
\be
 S = \frac{\pi}{2} \left[\sum_I a_I^2 + \sum_J b_J^2 \right]+  
        \frac{\beta}{4}(A+B) - \frac{3\beta^2}{32\pi}K.
\ee


\subsection{Eguchi-Hanson}
 \label{sec:eguchi}

A non-compact instanton which is a limiting case of the Taub-NUT solution
is the Eguchi-Hanson metric \cite{Eguchi78},
\begin{equation}
 \label{eqn:Eguchi-Hanson}
 ds^2 = (1-\frac{N^4}{r^4})(\frac{r}{8N})^2(d\tau + 4N\cos\theta d\phi)^2
 + (1-\frac{N^4}{r^4})^{-1}dr^2 + \frac{1}{4}r^2d\Omega^2.
\end{equation}
The instanton is regular if we consider the region $r \geq N$, and let 
$\tau$ have period $8\pi N$.  The boundary at infinity is $S^3/{\cal Z}_2$ and
hence the metric is ALE.
There is a bolt of area $\pi N^2$ at $r=N$, which gives rise to a Misner string
along the $z$-axis.

To calculate the Hamiltonian for the Eguchi-Hanson metric we use as a 
reference background  an 
orbifold obtained by identifying Euclidean flat space mod ${\cal Z}_2$.  
This has a nut at the orbifold point at the origin, with a Misner string lying
along the $z$-axis. 
The Hamiltonian at infinity vanishes, 
\be
 H_\infty = 0,
\ee
as does the Hamiltonian on the Misner string,
\begin{equation}
 H_{MS} = 0.
\end{equation}
We then find that the area of Misner string, when the area of the background
string has been subtracted, is simply minus the area of the bolt.  Hence the
action and entropy are both zero,
\begin{equation}
 I = 0 \spaceand S = 0.
\end{equation}
This is what one would expect, because Eguchi-Hanson has the 
the same supersymmetry as its reference background. It is only when the 
solution has less supersymmetry than the background that there is 
entropy.


\subsection{Five-Sphere}
\label{sec:FiveSphere}

Finally, to show that the expression we propose for the entropy, equation  
(\ref{eqn:entropy}), can be applied in more than four dimensions, consider the 
five-sphere of radius R,
\be
 ds^2 = R^2( d\chi^2 + \sin^2\chi ( d\eta^2 + \sin^2\eta(
  d\psi^2 + \sin^2\psi(d\theta^2 + \sin^2\theta d\phi^2 )))).
\ee
This can be regarded as a solution of a five-dimensional theory with 
cosmological constant $\Lambda=6/R^2$. 
If we consider dimensional reduction with respect to the $U(1)$ isometry
$\partial_\phi$, then the fixed point set is a  
three sphere of radius R. There are no Misner strings, so our formula  
gives an entropy equal to the area of the bolt, 
\be
 S = \frac{\pi^2 R^3}{2G}. 
\ee

However, one can choose a different $U(1)$ isometry, whose orbits are the Hopf 
fibration of the five sphere.  In this case, we want to write the metric as 
\be
 ds^2 = (d\tau + \omega_i dx^i)^2 + \frac{R^2}{4} \left[ d\sigma^2 + 
    \sin^2\frac{\sigma}{2}( \sigma_1^2 + \sigma^2_2 + \cos^2\frac{\sigma}{2}
      \sigma_3^2 )\right],
\ee 
where
\be
 \omega = \frac{R}{2}(-\cos^2\frac{\sigma}{2} \sigma_3 + \cos\theta d\phi),
\ee
the periodicity of $\tau$ is $2\pi R$, the range of $\sigma$ and $\theta$
is $[0,\pi]$ and the periodicities of $\psi$ and $\phi$ are $4\pi$ and $2\pi$ 
respectively. The isometry $\partial_\tau$ has no fixed points.  So the 
usual connection 
between entropy and fixed points does not apply.  The orbit space of 
the Hopf fibration is ${\cal C}P^2$ with the Kaluza-Klein two-form, 
$F=d\omega$, equal to the harmonic two-form on ${\cal C}P^2$. 
The one-form potential, $\omega$, has a Dirac string on 
the two-surface in the orbit space given by $\theta=0,\pi$.   
When promoted to the full spacetime, this becomes a 
three-dimensional Misner string of area 
\be
 {\cal A} = 4 \pi^2 R^3.  
\ee
Calculating the Hamiltonian surface term on the Misner string, we find 
\be
 H_{\rm MS} = \frac{\pi R^2}{4G}.
\ee
Hence, we see that the entropy is
\be
 S = \frac{\cal A}{4G} - \beta H_{\rm MS} =  \frac{\pi R^2}{2G}.
\ee
While this example is rather trivial, it does demonstrate that the
entropy formula (\ref{eqn:entropy}) can be extended to higher dimensions.


\section{Conclusions}
\label{sec:Conclusions}

There are three morals that  can be drawn from this work. The first is that 
gravitational entropy just depends on the Einstein-Hilbert action. It doesn't 
require supersymmetry, string theory, or p-branes.  Indeed, one can define 
entropy for the Taub-Bolt solution which does not admit a spin structure, at
least of the ordinary kind.  
The second moral is that entropy 
is a global quantity, like energy or angular momentum, and shouldn't be 
localized on the horizon. The various attempts to identify the microstates 
responsible for black hole entropy are in fact constructions of dual theories 
that live in separate spacetimes. 
The third moral is that entropy arises from a failure to foliate the Euclidean 
regime with a family of time surfaces.  In these situations the Hamiltonian
will not give a unitary evolution in time.  This raises the possibility of loss
of information and quantum coherence.

\section{Acknowledgments}

CJH acknowledges the financial support of the Association of Commonwealth
Universities and the Natural Sciences and Engineering Research Council of 
Canada.

\appendix

\section{Gauge Invariance of $H_\infty$}
\label{app:Infinity}

We are interested in making gauge transformations which shift the
Euclidean time coordinate,
\begin{equation}
 \label{eqn:gauge2}
 d\hat{\tau} = d\tau - 2\lambda_{,i} dx^i.
\end{equation}
Under this transformation the Hamiltonian variables transform as
\begin{eqnarray}
 \hat{N}^2 & = & \rho N^2,  \\
 \hat{N}_i & = & N_i + 2(N^2 + N_kN^k)\lambda_{,i}, \\
 \hat{N}^i & = & \rho(1+2\lambda_{,k}N^k) N^i + 2\rho N^2\lambda^{,i}, \\
 \hat{h}_{ij} & = & h_{ij} + 2N_{(i}\lambda_{,j)} + 4(N^2 + N_kN^k)\lambda_{,i}
   \lambda_{,j}, \\
 \hat{h}^{ij} & = & h^{ij} + \rho[2\lambda^2N^iN^j -
       4N^2\lambda^{,i}\lambda^{,j} - 2(1+2\lambda_{,k}N^k)N^{(i}\lambda^{,j)}],
\end{eqnarray}
where
\begin{equation}
 \rho = \frac{1}{2\lambda^2N^2 + (1+2\lambda_{,k}N^k)^2},
\end{equation}
and $\lambda^2 = \lambda_{,i}\lambda^{,i}$.  Indices for hatted terms are
raised and lowered with $\hat{h}_{ij}$, while those without are raised and
lowered by $h_{ij}$.
The total Hamiltonian is not invariant under such a transformation.
However, the Hamiltonian contribution at infinity will be shown to
be invariant for AF, ALF and ALE metrics.

The general asymptotic form of the AF metric is
\be
 ds^2 \sim \left(1-\frac{2M}{r}\right)d\tau^2  -
     \left(1+\frac{2M}{r}\right)[dr^2 +
     r^2d\Omega^2_2]
\ee
We can apply a general gauge transformation (\ref{eqn:gauge2}) to this,
where we asymptotically expand $\lambda$ as
\be
 \lambda \sim \lambda_0 + \frac{\lambda_1}{r} + {\cal O}(r^{-2}).
\ee
If we calculate the Hamiltonian after applying this gauge transformation, we
find that
\be
 \hat{H}_\infty = -r + M. 
\ee
In order calculate the background value, we need to scale flat space so that
the metrics agree of a surface of constant radius $R$.  The metric is
\be
 d\tilde{s}^2 = \left(1-\frac{2M}{R}\right)d\tau^2 +
         \left(1-\frac{2M}{R}\right)[dr^2 + r^2d\Omega^2_2].
\ee
Applying the gauge transformation, and then calculating the Hamiltonian yields
\be
 \hat{\tilde{H}}_\infty = -r.
\ee
Thus we see that the physical Hamiltonian is 
\be
 \hat{H}_\infty = M,
\ee
which is gauge invariant.

We now want to consider the value of the Hamiltonian at infinity for ALF
spaces.  The general asymptotic form of the ALF metric is
\be
 ds^2 \sim \left(1-\frac{2M}{r}\right)(d\tau + 2aN\cos\theta d\phi)^2  -
     \left(1-\frac{2M}{r}\right)[dr^2 + r^2d\Omega^2_2].
\ee
If we calculate the Hamiltonian after applying a gauge transformation then 
we find that, identical to the AF case,
\be
 \hat{H}_\infty = -r + M.
\ee
In order calculate the background value, we need the matched
ALF background metric,
\ba
 d\tilde{s}^2 & = & \left(1-\frac{2N}{r} -\frac{2(M-N)}{R}\right)
      (d\tau + 2aN\cos \theta d\phi)^2 + \nonumber \\
  & &  \left(1-\frac{2N}{r} + \frac{2(M-N)}{R} \right)[dr^2 + r^2d\Omega^2_2],
\ea
which has the gauge independent Hamiltonian,
\be
 \hat{\tilde{H}}_\infty = -r + N.
\ee
Thus we see that the physical Hamiltonian is gauge invariant, 
\be
 \hat{H}_\infty = M - N.
\ee

The general asymptotic form of the ALE metric is
\be
   ds^2 = \left(1+\frac{M}{r^4}\right) d{\cal E}_4^2 + {\cal O}(r^{-5}).
\ee
We note that the asymptotic background metric is simply the $M=0$ case
of the general metric, and hence the physical Hamiltonian is
\be
 H_\infty = H(M) - H(0).
\ee
If we calculate the Hamiltonian after applying the gauge transformation, then we
get a very complicated function of $M$, $R$ and $\lambda$.  However, if we
differentiate with respect to $M$, we find that
\be
 \frac{\partial \hat{H}_\infty}{\partial M}  = {\cal O}(r^{-2}).
\ee
Thus, the background subtraction will cancel the Hamiltonian up to ${\cal O}(r^{-2})$, and hence
\be
 \hat{H}_\infty = 0,
\ee
which is obviously gauge invariant.


\begin{thebibliography}{                                     }

\bibitem{Christodoulou70} 
 D. Christodoulou,
 ``Reversible and Irreversible Transformations in Black Hole Physics'',
 {\it Phys. Rev. Lett.} {\bf 25}, 1596 (1970).

\bibitem{Hawking71}
 S.W. Hawking,
 ``Gravitational Radiation from Colliding Black Holes'',
 {\it Phys. Rev. Lett.} {\bf 26}, 1344 (1971).

\bibitem{BardeenCarterHawking73}
 J.M. Bardeen, B. Carter and S.W. Hawking,
 ``The Four Laws of Black Hole Mechanics'',
 {\it Comm. Math. Phys.} {\bf 31}, 161 (1970).

\bibitem{Bekenstein73}
 J.D. Bekenstein,
 ``Black Holes and Entropy'',
 {\it Phys. Rev.} {\bf D7}, 2333 (1973).

\bibitem{Hawking74}
 S.W. Hawking,
 ``Black Hole Explosions'',
 {\it Nature} {\bf 248}, 30 (1974).

\bibitem{Hawking75}
 S.W. Hawking,
 ``Particle Creation by Black Holes'',
 {\it Comm. Math. Phys.} {\bf 43}, 199 (1975).

\bibitem{StromingerVafa96}
 A. Strominger and C. Vafa,
 ``Microscopic Origin of the Bekenstein-Hawking Entropy'',
 {\it Phys.Lett.} {\bf B379}, 99 (1996).

\bibitem{AntideSitter}
 J.M. Maldacena,
 ``The Large N Limit of Superconformal Field Theories and Supergravity'',
 hep-th/9711200.
 E. Witten,
 ``Anti De Sitter Space And Holography'',
 hep-th/9802150.
 S.S. Gubser, I.R. Klebanov and A.M. Polyakov, 
 ``Gauge Theory Correlators from Non-Critical String Theory'',
 {\it Phys. Lett.} {\bf B428}, 105 (1998).
 S.S. Gubser, I.R. Klebanov and A.W. Peet, 
 ``Entropy and Temperature of Black 3-branes'', 
 {\it Phys. Rev.} {\bf D54}, 3915 (1996).
 I.R. Klebanov and A. A. Tseytlin, 
 ``Entropy of Near-extremal Black p-branes'',
 {\it Nucl. Phys.} {\bf B475}, 164 (1996).
 
\bibitem{ArnowittDeserMisner62} 
 R. Arnowitt, S. Deser, and C. Misner, 
 in {\it Gravitation:  An Introduction to Current Research}, ed. L. Witten,
 Wiley, New York (1962).

\bibitem{HawkingHorowitz96}
 S.W. Hawking and G.T. Horowitz,
 ``The Gravitational Hamiltonian, Action, Entropy, and Surface Terms'',
 {\it Class. Quant. Grav.} {\bf 13}, 1487 (1996).

\bibitem{HawkingHunter96}
 S.W. Hawking and C.J. Hunter,
 ``The Gravitational Hamiltonian in the Presence of Non-Orthogonal Boundaries'',
 {\it Class. Quant. Grav.} {\bf 13}, 2735 (1996).

\bibitem{GibbonsHawking79}
 G.W. Gibbons and S.W. Hawking,
 ``Classification of Gravitational Instanton Symmetries'',
 {\it Comm. Math. Phys.} {\bf 66}, 291 (1979).

\bibitem{Dowker96}
 H.F. Dowker, J.P. Gauntlett, G.W. Gibbons and G.T. Horowitz,
 ``Nucleation of $P$-Branes and Fundamental Strings'',
 {\it Phys.Rev.} {\bf D53} 7115 (1996).

\bibitem{Hunter98}
 C.J. Hunter,
 ``The Action of Instantons with Nut Charge'',
 gr-qc/9807010.

\bibitem{Misner63}
 C.W. Misner,
 ``The Flatter Regions of Newman, Unti, and Tamburino's Generalized 
   Schwarzschild Space'',
 {\it Journal of Math. Phys.} {\bf 4}, 924 (1963). 

\bibitem{Hawking77}
 S.W. Hawking,
 ``Gravitational Instantons'',
 {\it Phys. Lett.} {\bf 60A}, 81 (1977).

\bibitem{Page78}
 D.N. Page,
``Taub-NUT Instanton with an Horizon'',
 {\it Phys. Lett.} {\bf 78B}, 249 (1978).

\bibitem{Hawking99}
 S.W. Hawking, C.J. Hunter and D.N. Page,
 ``NUT Charge, Anti-de Sitter space and Entropy'',
 paper in preparation.

\bibitem{Whitt85}
 B. Whitt,
 ``Israel-Wilson Metrics'',
 {\it Annals of Physics} {\bf 161}, 241 (1985).

\bibitem{Yuille87}
 A.L. Yuille,
 ``Israel-Wilson metrics in the Euclidean Regime'',
 {\it Class. Quant. Grav.} {\bf 4}, 1409 (1987).

\bibitem{GibbonsHawking78}
 G.W. Gibbons and S.W. Hawking
 ``Gravitational Multi-Instantons'',
 {\it Phys. Lett.} {\bf 78B}, 430 (1978).

\bibitem{Eguchi78}
 T. Eguchi and A.J. Hanson,
 ``Asymptotically Flat Self-Dual Solutions to Euclidean Gravity'',
 {\it Phys. Lett.} {\bf 74B}, 249 (1978).



\end{thebibliography}
\end{document}